\documentclass[final,numberedheadings]{aipproc}
\layoutstyle{6x9}
\begin{document}
\title{Recent Developments in Lattice QCD}
\author{Gunnar S.\ Bali}{
  address={Department of Physics \& Astronomy, The University of Glasgow, Glasgow G12 8QQ, UK}
}
\begin{abstract}
Recent trends in lattice QCD calculations
and latest results are reviewed, with particular emphasis on
multiquark-states and the nucleon structure.
\end{abstract}
\maketitle
\section{Introduction}
While the simplicity and elegance of QCD is very appealing theoretically,
the phenomenological observations of spontaneous chiral symmetry breaking
and of the confinement of colour charges turned it
into a major calculational nightmare: it took almost twenty years after
the discovery of asymptotic freedom in 1973 to convincingly
demonstrate that the QCD Lagrangian indeed implies these highly
non-trivial collective phenomena. This was done by 
means of numerical simulation.

In such lattice simulations, QCD is regularised by introducing a lattice
cut-off $a$. Subsequently, the $n_f+1$ parameters of the theory,
i.e.\ QCD coupling and quark masses are matched to reproduce $n_f+1$
hadron masses. Everything else is a prediction and in this sense lattice QCD is
a {\em first principles} approach. The confinement of colour
implies that finite size effects are usually tiny, as long as
the spatial box extent $La\gg m_{\pi}^{-1}$.
We are also fortunate to find that lattice spacings
$a^{-1}=1$ -- 4~GeV are sufficiently small
to allow for controlled continuum limit extrapolations,
$a\rightarrow 0$. This means that
$L\ll 100$ is sufficient, which makes QCD tractable on computers.

This need not be so. If for instance we were to attempt a brute
force calculation of the deuteron binding energy of about 2~MeV then
resolving the large scale difference with respect to the nucleon mass of almost
a factor of 50 would require both, a huge volume and incredible statistical 
accuracy. Of course
such scale hierarchies indicate that direct calculations should be
avoidable as they lend themselves to an effective field theory
treatment.

We will discuss progress on
computational problems in Sec.~\ref{sec:cost},
before summarising recent lattice studies of multiquark hadron spectroscopy
in Sec.~\ref{sec:multi} and of the nucleon structure in Sec.~\ref{sec:nucleon}.
\section{Challenges in lattice calculations}
\label{sec:cost}
On a lattice with $V=L^3T$ points, the lattice Dirac operator is a
huge matrix of dimension $12V$. The inversion of this operator
represents the major
computational task of lattice QCD and this makes
simulations incorporating sea quarks notoriously expensive.
The algorithmic
cost explodes
with small $\pi$ masses, $\propto1/(m_{\pi}a)^{\simeq 3}$. A smaller
$m_{\pi}$ also requires a larger spatial lattice volume and the 
scaling behaviour, keeping
$m_{\pi}L$ fixed, is even worse: $\propto 1/(m_{\pi}a)^{\simeq 7}$.

\begin{figure}
  \includegraphics[height=.32\textheight]{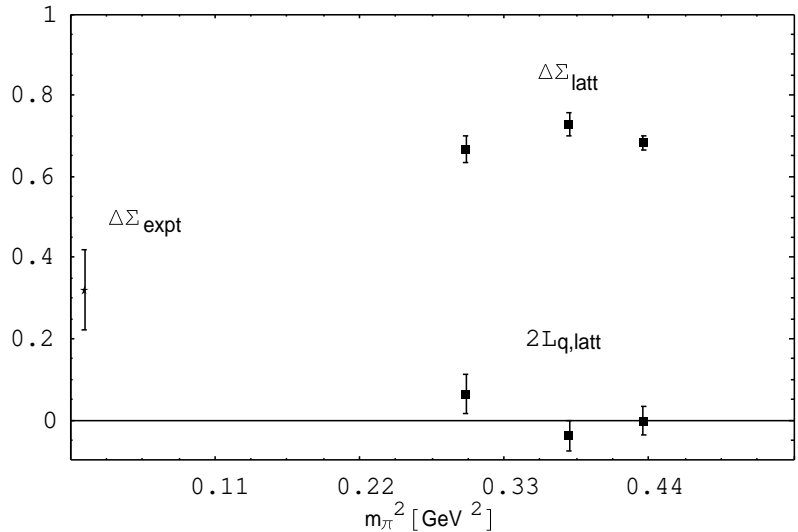}
  \caption{The fraction of the nucleon spin carried by the quarks~\cite{Negele:2004iu}.}
\label{fig:spin}
\end{figure}
In Fig.~\ref{fig:spin}
we display recent results obtained by the LHP and SESAM
Collaborations~\cite{Negele:2004iu}
on the quark contribution $\Delta\Sigma$ to the proton spin,
in the $\overline{MS}$ scheme at a scale $\mu=2$~GeV, for $n_f=2$.
The normalisation is such that $\frac12=\frac12\Delta\Sigma+L_q+J_g$,
where $L_q$ is the contribution from the quark angular momentum and
$J_g$ from the gluons. Similar results have been obtained by
the QCDSF Collaboration~\cite{Gockeler:2003jf}.
In these simulations, $m_{\pi}>550$~MeV.
We took the liberty to convert the mass scale into units of our choice.
Obviously, for infinite quark masses we expect $\Delta\Sigma=1$. It
is therefore not surprising that the experimental value is overestimated.
However, it is clear from the figure that smaller quark masses
are absolutely essential to allow for a meaningful chiral extrapolation.
Fortunately, with the advent of new Fermion
formulations~\cite{Neuberger:1997fp} that
respect an exact lattice chiral symmetry,
a reduction of the quark mass towards
values $m_{\pi}\approx 180$~MeV has become possible
recently~\cite{Dong:2003zf},
albeit so far only in the quenched approximation.

\begin{figure}
\includegraphics[height=.28\textheight]{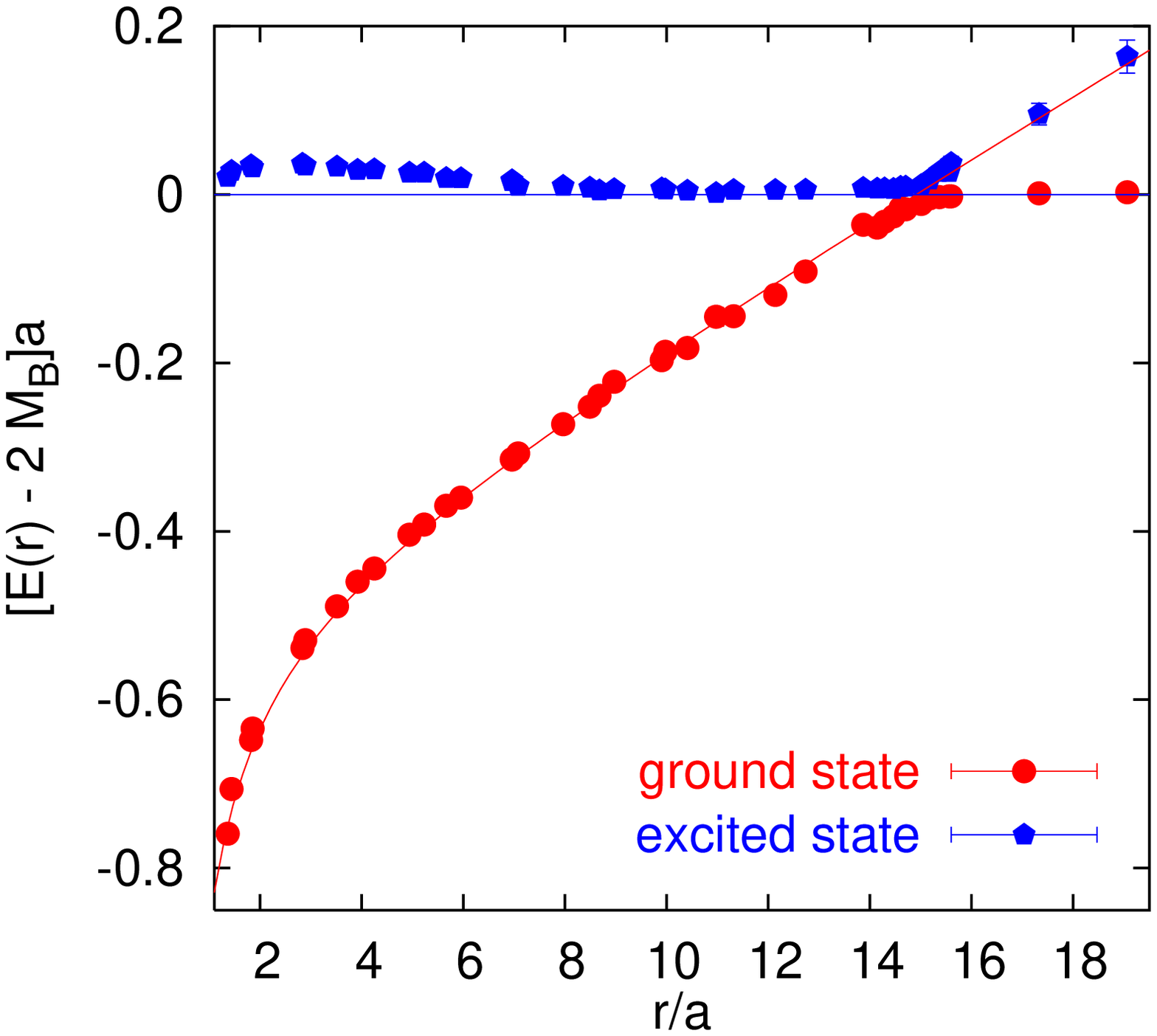}~~~
\includegraphics[height=.28\textheight]{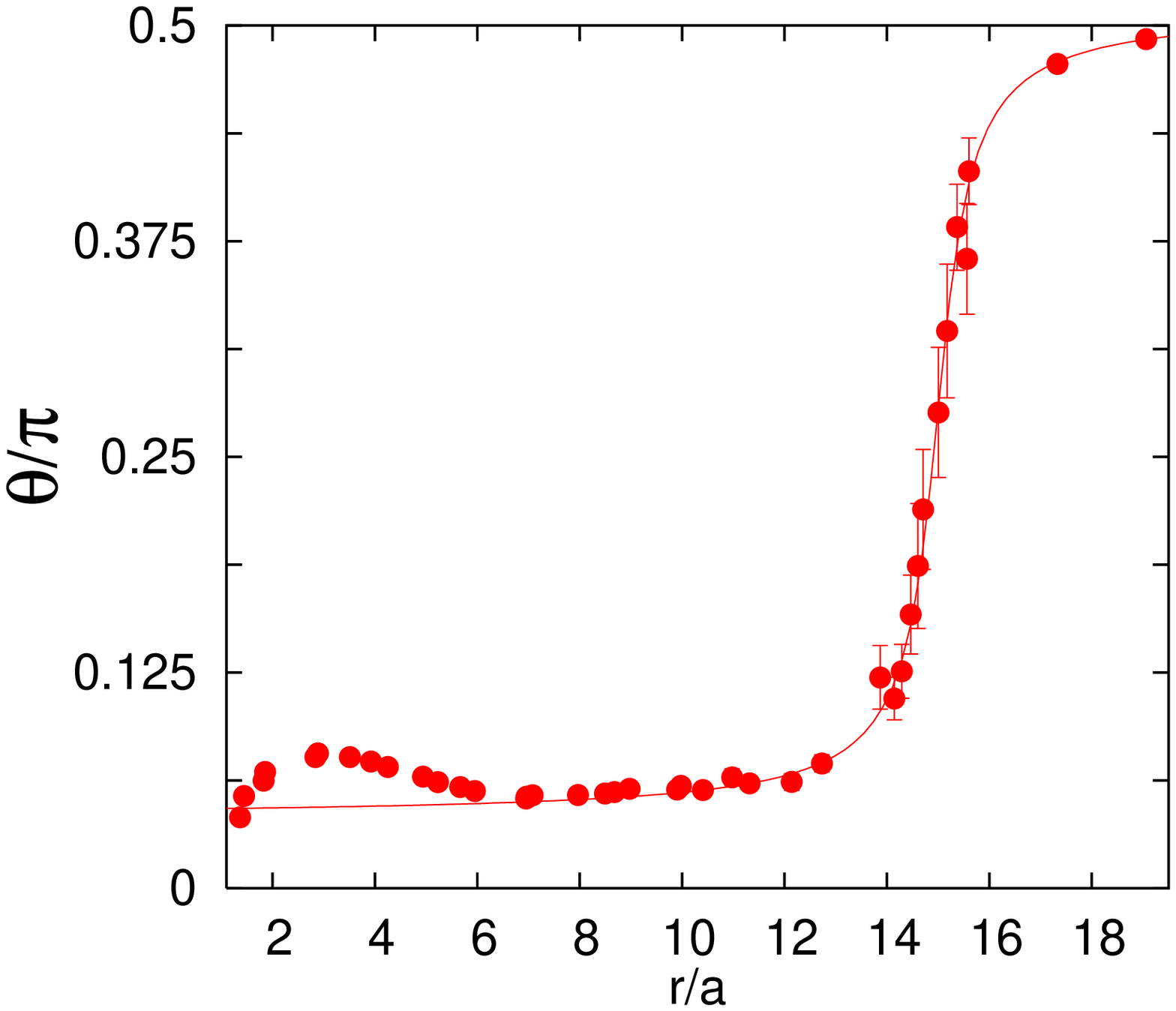}
\caption{The lowest two energy levels in the $Q{\overline Q}$ system
with sea quarks. On the right hand side the $\overline{B}B$ content
of the ground state is shown, in terms of a mixing angle.
The results apply to $n_f=2$, $m_q\approx
m_s$ and $a\approx 0.085$~fm~\cite{Bali:2004pb}.}
\label{fig:break}
\end{figure}
In many lattice calculations it is sufficient to calculate quark propagators
that originate from a fixed source point. 
In these cases only one
column of the inverse Dirac matrix needs to be calculated,
na\"{\i}vely reducing the
effort by a factor $V$.
In some cases diagrams with disconnected quark lines are needed.
Examples are the physics
of flavour singlet mesons, strong decays as well as
parton distributions (PDs). In the latter case the complication can
be avoided by
assuming $SU(2)$ isospin symmetry and
only calculating differences between $u$ and $d$ quark distributions.
Disconnected quark lines require all-to-all propagators and hence 
a complete inversion of the Dirac matrix appears necessary.
This turns out to be prohibitively expensive in terms of memory
and computer time. Fortunately,
sophisticated noise reduced stochastic estimator
techniques have been developed over the past few years and as a result
tremendous progress was achieved. One such benchmark is the
QCD string breaking
problem, $Q(r)\overline{Q}(0)\leftrightarrow\overline{B}(r)B(0)$, where
$B=\overline{Q}q$ and $Q$ is a static quark, which represents
one of the cleanest examples of a strong decay.
Within both, the transition matrix element as well as the $\overline{B}B$
state all-to-all propagators are required.
In Fig.~\ref{fig:break} we display the result of a recent SESAM Collaboration
study~\cite{Bali:2004pb}.
An extrapolation to physical light quark masses yields a string breaking
distance $r_c\approx 1.16$~fm.
The gap between
the two states in the string breaking region is
$\Delta E\approx 50$~MeV and we are
able to resolve this with a resolution of 10 standard deviations!

In conclusion, all the long standing killers $m\ll m_s$, $n_f>0$ and
all-to-all propagators have been successfully tackled. However, we are still
a few years away from overcoming two of them
at the same time and possibly
almost a decade separates us from simulations of flavour singlet
diagrams with realistically light sea quarks. In most cases
not all these ingredients are required simultaneously.
While bigger computers are always needed, most of the recent progress
would have been impossible without novel methods.
The gain factor from
faster computers was almost 5,000 over the past 15 years.
The factor from theoretical and algorithmic advances is harder to quantify.
Two such examples have been
mentioned: chiral lattice Fermions and all-to-all propagator techniques.
\section{Pentaquarks and tetraquarks}
\label{sec:multi}
QCD goes beyond the quark model and hence
hadronic states that do not fit into a na\"{\i}ve quark model of
$q\bar{q}$ mesons and $qqq$ baryons are of particular interest.
Many of the observed hadrons will contain considerable non-quark model
components, in particular within the scalar sector. Obviously,
quantum numbers that are incomprehensible with a quark model meson
or baryon interpretation provide us with the most 
clean-cut distinction. Such examples do exist in the Review of Particle
Properties, namely the $J^{PC}=1^{-+}$ mesons $\pi_1(1400)$
and $\pi_1(1600)$. The minimal configuration required to obtain a vector
state with positive charge either consists of two quarks and two antiquarks
(tetraquark/molecule) or of quark, antiquark and a gluonic excitation
(hybrid meson). Unfortunately, these resonances are rather broad
with a width 
$\Gamma\approx$~300~MeV
which might be one of the reasons why they are often ignored.
However, the ratio $\Gamma/m$ is very much the same as for the
(quark model) $\rho(770)$ meson.

Another clear indication of an $n_{\rm quark}>3$
nature would for instance
be a baryonic state with strangeness $S=+1$. The minimal quark configuration
in this case consists of five quarks (pentaquark): $uudd\bar{s}$.
It is no secret that over the past two years several experiments
have presented evidence of a very narrow
$\Theta^+(1530)$ resonance, with decay
$\Theta^+\rightarrow pK^0$ and $\Theta^+\rightarrow nK^+$.
The parity has not yet been established. However the mass is
about 100~MeV above the $KN$ threshold and for $J^P=1/2^-$
an $S$-wave decay is possible, which is difficult to reconcile with
a width $\Gamma \ll 10$~MeV. For $1/2^+$ a $P$-wave is required, still
a bit puzzling but less so. As the main decay channel does not require
sea quarks one might hope to gain some
insight from quenched lattice simulations
and several attempts have been
made~\cite{Csikor:2003ng,Sasaki:2003gi,Chiu:2004gg,Mathur:2004jr,Ishii:2004qe,Alexandrou:2004ws}.

Two groups~\cite{Sasaki:2003gi,Chiu:2004gg} also investigated charmed
pentaquarks and two
studies~\cite{Csikor:2003ng,Mathur:2004jr} incorporated the $I=1$ sector,
in addition to $I=0$. Two groups~\cite{Chiu:2004gg,Mathur:2004jr}
employed chiral overlap Fermions while the others used conventional
Wilson-type lattice quarks. Only the 
Kentucky group~\cite{Mathur:2004jr} went down to $m_{\pi}\approx 180$~MeV
while all other $\pi$ masses were larger than 400~MeV. The
Budapest-Wuppertal group~\cite{Csikor:2003ng} varied the lattice spacing and
attempted a continuum-limit extrapolation.
In all studies the negative parity mass came out to be lighter than the
positive parity one, which is expected in the heavy quark limit.

There are two crucial questions to be asked: what happens when
realistic light quark masses are approached? Do we see
resonant or scattering states?
Resolving a resonance sitting on top of a tower of
$KN$ scattering states with different relative momenta appears rather
hopeless at first. However, there are two discovery tools available:
variation of the lattice volume and of the creation operator.
By varying the volume one will change the spectrum of $KN$ scattering states
as well as the coupling of a given operator to $KN$ (the spectral weight).
If the pentaquark really was such a narrow resonance
as some experiments suggest then maybe a lattice operator can be constructed
that has a large overlap with this state but only a very small coupling
to $KN$. For the $1/2^+$ state which can
only decay into a $P$-wave, the mass of the scattering
state will depend on the lattice size since the smallest possible non-vanishing
lattice momentum is $\pi/(aL)$. For $1/2^-$ the volume dependence of the
lowest scattering state mass will be weak, however, the scaling of the
spectral weight with the volume can still provide us with a hint.

\begin{figure}
\includegraphics[height=.23\textheight]{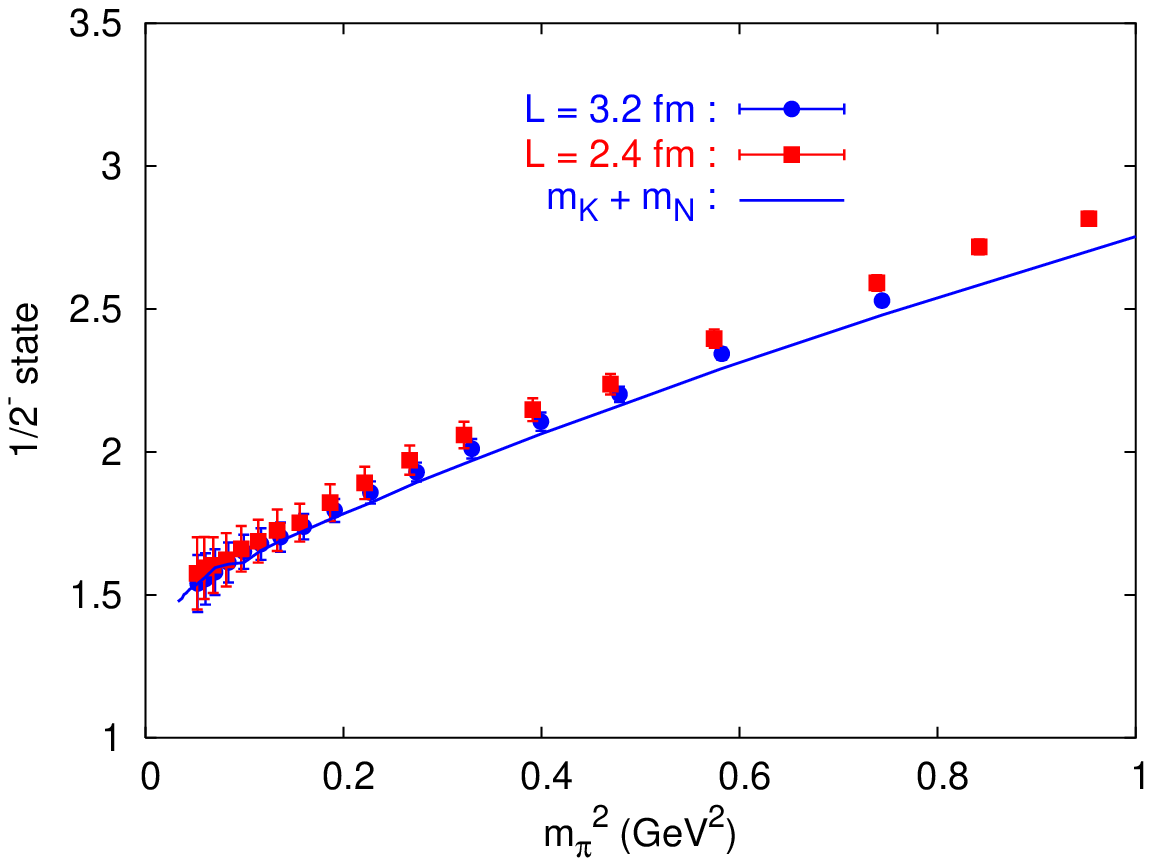}~~~
\includegraphics[height=.23\textheight]{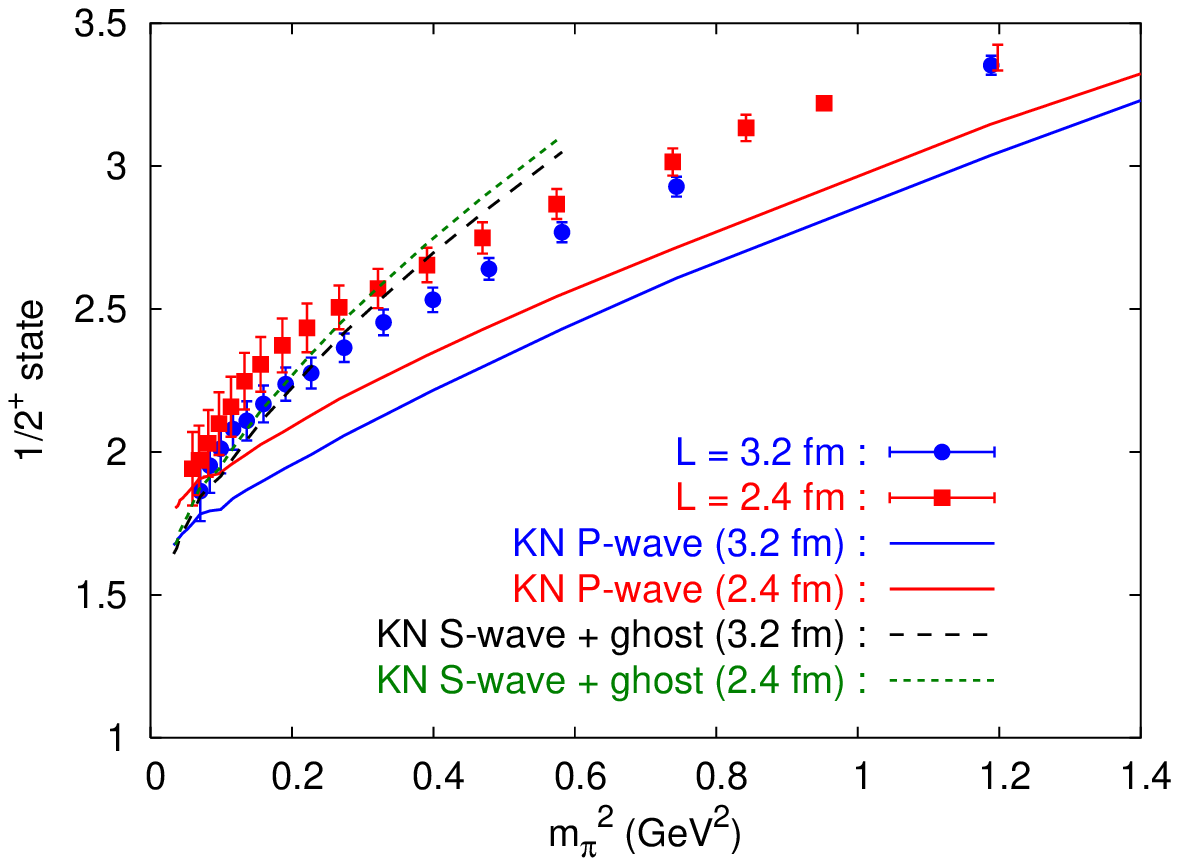}
\caption{The $\pi$ mass dependence of $I=0$
$uudd\bar{s}$ ``pentaquark'' masses~\cite{Mathur:2004jr}.}
\label{fig:penta}
\end{figure}
It turns out that the situation on the lattice is at least as ambiguous as
the one encountered in experiment. To demonstrate this we display some
Kentucky results~\cite{Mathur:2004jr} in Fig.~\ref{fig:penta}.
It appears that the $1/2^-$ state dominantly couples to an
$S$-wave $KN$. The $1/2^+$ however displays the qualitative volume dependence
of a $P$-wave but does not share its mass.
Interpretation as a $P$-wave plus some linear
momentum is a likely possibility. At very small $m_{\pi}$ the situation
becomes further complicated by the fact that there is no axial anomaly in the
quenched approximation. Hence the flavour singlet $\eta'$ is degenerate
with the $\pi$. As this contribution comes in with a negative spectral weight,
it is sometimes labelled as a ``ghost''. The $1/2^+$ state can contain
such a $KN\eta'$ $S$-wave (dashed lines). While the $1/2^-$ state
becomes indistinguishable form a $KN$, the $P=1/2^+$ {\em might}
contain a resonant component. In order to arrive at more definite conclusions
a variation of the creation operator as well as of the volume appears
necessary, which is a very ambitious project.
Nonetheless, the lattice results obtained so far are quite instructive.

Lattice studies of diquark interactions 
in a simplified, more controlled environment represent
an alternative strategy to the brute force simulation of unstable states.
A baryon with one static and two light quarks constitutes one such arena.
One can of course also investigate multiquark interactions in the nonrelativistic
limit of infinitely heavy quark masses. Such tetra- and penta-quark potentials
have been studied recently by two
groups~\cite{Alexandrou:2004ws,Alexandrou:2004ak,Okiharu:2004wy}
and the results
might provide model builders with some insight. However, it is not
clear how to relate these findings to the light quark limit in which
chiral symmetry appears to play a bigger r\^ole than instantaneous
confining forces.

There exist quite a few narrow resonances very close to strong decay thresholds
like the $\Lambda(1405)$, the recently discovered $X(3872)$ charmonium state
and the $a_0/f_0(980)$ system. It is very conceivable that such states
contain a sizable multiquark component. The question then arises
if these constitute would-be quark model states or if these are
true molecules/multiquark-states, that appear {\em in addition}
to the quark model
spectrum. A fantastic arena to
address this was provided by the recently discovered (probably scalar)
$D_s^*(2317)$ and (probably axialvector) $D_s^*(2457)$ states.
First lattice studies~\cite{Bali:2003jv,Dougall:2003hv}
have been performed,
with somewhat contradictory interpretations of very compatible results.
One might hope that a similar effort will be dedicated on
the comparatively cleaner and easier question of tetraquarks as
has been on the pentaquark studies.
\section{The nucleon: form and structure}
\label{sec:nucleon}
The obvious strength of lattice studies lies in hadron spectroscopy.
Calculations of hadronic matrix elements are more involved.
However, once light sea quarks are included, many states become
unstable and it turns out that calculating internal properties of stable
particles is easier than resolving the strong decay dynamics of
resonances. This should also be clear from the pentaquark discussion above.
In view of this, the phenomenologically most exciting lattice input
to be expected in the next few years should be calculations of the
hadron structure.

Quite a few results on moments of generalised parton distributions
(GPDs), most notably
from the QCDSF~\cite{Gockeler:2004vx}, SESAM and
LHP Collaborations~\cite{Hagler:2003jd}, exist and the lattice
method is already very competitive in this area
which is experimentally very challenging, in particular when it comes to
transversity distributions.
These new simulations build upon the methods that have
been developed and successfully used in the context
of calculations of unpolarised and polarised PDs. Possibly the
biggest systematic uncertainty in these past studies was the
chiral extrapolation, down to physical $m_{\pi}$.

\begin{figure}
\includegraphics[height=.26\textheight]{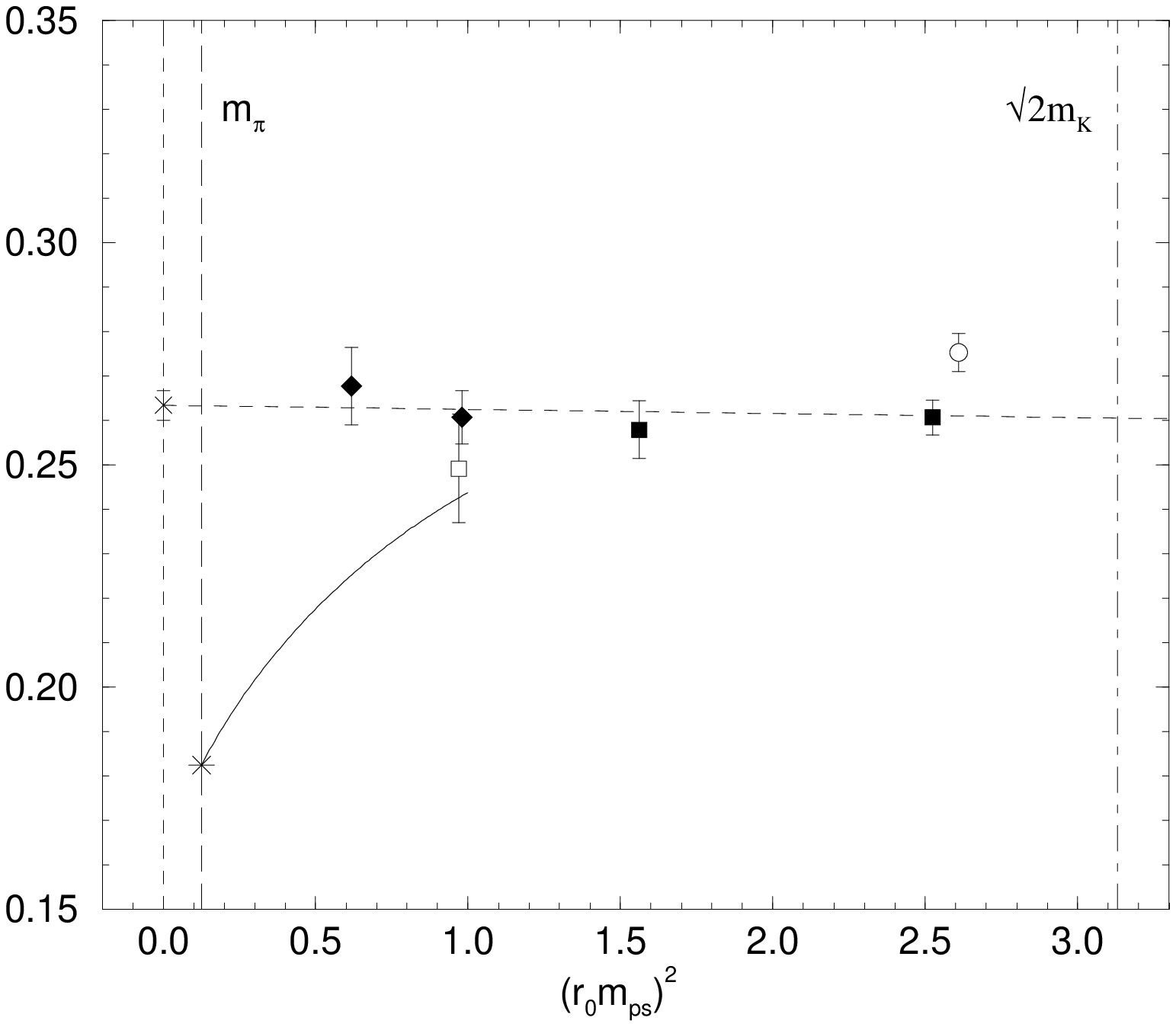}~~~~
\includegraphics[height=.26\textheight]{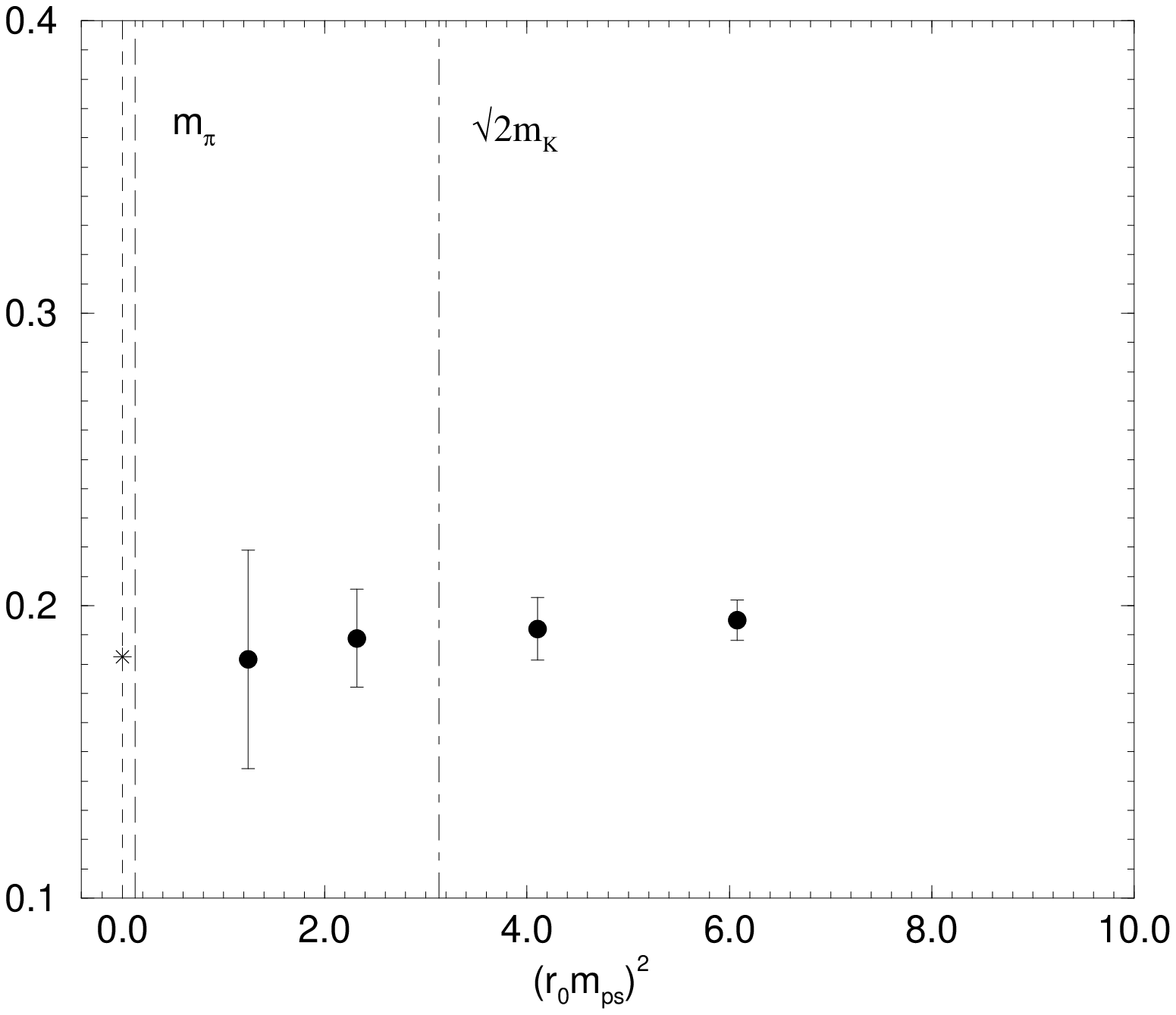}
\caption{The moment $\langle x\rangle_{u-d}$ of the proton quark distribution
with Wilson Fermions (left) and with overlap Fermions (right)~\cite{Bakeyev:2003gj},
in the $\overline{MS}$ scheme at 2~GeV. The units are
$r_0^{-1}\approx 400$~MeV.}
\label{fig:pd}
\end{figure}
In particular the lowest moment of
the unpolarised distribution,
\begin{equation}
\langle x\rangle_{u-d}=\langle x\rangle_u-\langle x\rangle_d,\quad
\langle x^n\rangle_q=\int_0^1\!dx\,x^n\left[q(x)-(-)^n\bar{q}(x)\right],
\end{equation}
turned out to be overestimated by almost 60~\% if linearly extrapolated
in $m_{\pi}^2$, both without and with sea
quarks~\cite{Dolgov:2002zm,Gockeler:2004wp}. Chiral perturbation theory
predicts deviations from such a linear behaviour and there
is some room for inventive extrapolations~\cite{Dolgov:2002zm}. However,
even at $m_{\pi}\approx 300$~MeV there is no sign of a bending towards the
chiral limit~\cite{Bakeyev:2003gj} as can be seen from Fig.~\ref{fig:pd}.
In the right part of the figure
the study is repeated with chiral overlap
Fermions which protect against mixing with lower dimensional lattice operators,
down to $m_{\pi}\approx 450$~MeV. In this case no non-perturbative
renormalisation to an intermediate MOM scheme
is available as yet.
However, in the Wilson case the difference between purely perturbative
and (partly) non-perturbative renormalisation was negligible.
This is potentially exciting news which probably will receive clarification
during the coming year.

Finally, there has also been progress in resolving the momentum dependence
of electromagnetic $\gamma^* N\rightarrow\Delta$ transition form
factors in a quenched study~\cite{Alexandrou:2004xn},
in the region $0.1\,\mbox{GeV}^2<Q^2<
1.4\,\mbox{GeV}^2$. The magnetic dipole form factor is significantly
overestimated at large $Q^2$, due to the unrealistically small
charge radii of $\Delta$
and $N$. One might hope that such effects cancel in part from
form factor ratios. $R_{EM}=G_{E2}/G_{M1}$
is fairly constant at -0.02(1), once extrapolated to the chiral limit.
In contrast, $R_{CM}=G_{C2}/G_{M1}$ decreases monotonously from 
-0.01(1) at 0.1~GeV${}^2$ down to -0.09(3) at $Q^2>1\,\,\mbox{GeV}^2$.
This behaviour is
in good agreement with $Q^2>0.4\,\,\mbox{GeV}^2$
CLAS data~\cite{Joo:2001tw} while it
is hard to reconcile with the OOPS point~\cite{Sparveris:2004jn},
$R_{SM}=[-6.1\pm 0.2\pm 0.5]\,\%$ at $Q^2\approx 0.13\,\mbox{GeV}^2$.
\section{Conclusion}
Lattice calculations have made a lot of progress recently. Many studies now
include sea quarks. Within the quenched approximation, light quark masses close
to the physical limit have been realised and the lattice provides
a powerful tool for exploring the validity range of chiral expansions.
There has also been tremendous progress in the calculation of diagrams
with disconnected quark lines which are for instance
needed to understand strong decay
processes and flavour singlet contributions.
A lot of work still needs to be done.
I was only able to present a tiny selection
of latest lattice results and my apologies go to all
those whose work I failed to mention.
\begin{theacknowledgments}
I thank the organisers of EFB 19.
This work is supported by the
EC Hadron Physics I3 Contract No.\ RII3-CT-2004-506078,
by a PPARC Advanced
Fellowship (grant PPA/A/S/2000/00271) as well as by PPARC grant
PPA/G/0/2002/0463. 
\end{theacknowledgments}


\begin{thebibliography}{99}
\bibitem{Negele:2004iu}
J.~W.~Negele {\it et al.},
\emph{Nucl.\ Phys.\ Proc.\ Suppl.},  {\bf 128}, 170 (2004)
[arXiv:hep-lat/0404005].
\bibitem{Gockeler:2003jf}
M.~G\"ockeler {\em et al.}\
                  [QCDSF Collab.],
\emph{Phys.\ Rev.\ Lett.}, {\bf 92}, 042002 (2004)
[arXiv:hep-ph/0304249].
\bibitem{Neuberger:1997fp}
H.~Neuberger,
\emph {Phys.\ Lett.\ B}, {\bf 417}, 141 (1998)
[arXiv:hep-lat/9707022] and references therein.
\bibitem{Dong:2003zf}
S.~J.~Dong {\em et al.},
arXiv:hep-ph/0306199.
\bibitem{Bali:2004pb}
G.\,S.\ Bali, T.\ D\"ussel, T.\ Lippert, H.\ Neff, Z.\ Prkacin and K.\ Schilling,
[SESAM Collab.]
arXiv:hep-lat/0409137 and in preparation.
\bibitem{Csikor:2003ng}
F.~Csikor, Z.~Fodor, S.~D.~Katz and T.~G.~Kovacs,
\emph{JHEP} {\bf 0311}, 070 (2003)
[arXiv:hep-lat/0309090].
\bibitem{Sasaki:2003gi}
S.~Sasaki,
\emph{Phys.\ Rev.\ Lett.}\  {\bf 93}, 152001 (2004)
[arXiv:hep-lat/0310014].
\bibitem{Chiu:2004gg}
T.~W.~Chiu and T.~H.~Hsieh,
arXiv:hep-ph/0403020.
\bibitem{Mathur:2004jr}
N.~Mathur {\it et al.},
arXiv:hep-ph/0406196.
\bibitem{Ishii:2004qe}
N.~Ishii, T.~Doi, H.~Iida, M.~Oka, F.~Okiharu and H.~Suganuma,
arXiv:hep-lat/0408030.
\bibitem{Alexandrou:2004ws}
C.~Alexandrou, G.~Koutsou and A.~Tsapalis,
arXiv:hep-lat/0409065.
\bibitem{Alexandrou:2004ak}
C.~Alexandrou and G.~Koutsou,
arXiv:hep-lat/0407005.
\bibitem{Okiharu:2004wy}
F.~Okiharu, H.~Suganuma and T.~T.~Takahashi,
arXiv:hep-lat/0407001.
\bibitem{Bali:2003jv}
G.~S.~Bali,
\emph{Phys.\ Rev.\ D} {\bf 68}, 071501 (2003)
[arXiv:hep-ph/0305209].
\bibitem{Dougall:2003hv}
A.~Dougall {\em et al.}\
[UKQCD Collab.],
\emph{Phys.\ Lett.\ B} {\bf 569}, 41 (2003)
[arXiv:hep-lat/0307001].
\bibitem{Gockeler:2004vx}
M.~G\"ockeler {\it et al.}  [QCDSF Collab.],
arXiv:hep-lat/0409162;
\emph{Phys.\ Rev.\ Lett.}\  {\bf 92}, 042002 (2004)
[arXiv:hep-ph/0304249].
\bibitem{Hagler:2003jd}
P.~H\"agler {\em et al.}\
                  [LHP and SESAM Collab.],
\emph{Phys.\ Rev.\ D} {\bf 68}, 034505 (2003)
[arXiv:hep-lat/0304018];
{\em Phys.\ Rev.\ Lett.}\  {\bf 93}, 112001 (2004)
[arXiv:hep-lat/0312014].

\bibitem{Dolgov:2002zm}
D.~Dolgov {\it et al.}  [LHP and SESAM Collab.],
{\em Phys.\ Rev.\ D} {\bf 66}, 034506 (2002)
[arXiv:hep-lat/0201021].
\bibitem{Gockeler:2004wp}
M.~G\"ockeler, R.~Horsley, D.~Pleiter, P.~E.~L.~Rakow and G.~Schierholz  [QCDSF
                  Collab.],
arXiv:hep-ph/0410187.
\bibitem{Bakeyev:2003gj}
T.~Bakeyev {\it et al.}  [QCDSF and UKQCD Collab.],
\emph{Nucl.\ Phys.\ Proc.\ Suppl.}\  {\bf 128}, 82 (2004)
[arXiv:hep-lat/0311017];
M.~G\"urtler {\it et al.}, [QCDSF Collab.],
arXiv:hep-lat/0409164.
\bibitem{Alexandrou:2004xn}
C.~Alexandrou, P.~de Forcrand, H.~Neff, J.~W.~Negele, W.~Schroers and A.~Tsapalis,
arXiv:hep-lat/0409122.
\bibitem{Joo:2001tw}
K.~Joo {\it et al.}  [CLAS Collab.],
\emph{Phys.\ Rev.\ Lett.}\  {\bf 88}, 122001 (2002)
[arXiv:hep-ex/0110007].
\bibitem{Sparveris:2004jn}
N.~F.~Sparveris {\it et al.}  [OOPS Collab.],
arXiv:nucl-ex/0408003.
\end{thebibliography}
\end{document}